\documentclass[dvips,a4paper,onecolumn,11pt]{article}

\usepackage{graphicx}
\usepackage{amsmath}
\usepackage{amscd}
\usepackage{amssymb}
\usepackage{color}
\usepackage{subfigure}
\usepackage{supertabular}
\usepackage{array}
\usepackage{textcomp}
\usepackage{natbib}
\usepackage[in]{fullpage}
\usepackage[
bookmarks=false,
pdfauthor = {Laxalde~et~al.},
pdftitle={Qualitative Analysis of Forced Response of Blisks With Friction Ring Dampers},
pdfkeywords={Bladed-disk, friction damping, nonlinear dynamics, HBM}
]{hyperref}

\graphicspath{{EPS/}}

\newcommand{\vect}[1]{\mathbf{#1}}	% vector's format
\newcommand{\mat}[1]{\mathbf{#1}}	% matrices format 

\begin{document}

% \begin{frontmatter}
\noindent
\begin{flushleft}
   Journal home page: http://www.sciencedirect.com/science/journal/09977538 \\
   \vspace{1em}

   Qualitative analysis of forced response of blisks with friction ring dampers\\
   European Journal of Mechanics - A/Solids, Volume 26, Issue 4, July-August 2007, Pages 676-687\\
   D. Laxalde, F. Thouverez, J.-J. Sinou and J.-P. Lombard
\end{flushleft}
   \vspace{1em}
% 	\journal{European Journal of Mechanics}
%     \volume{26, Issue 4}
%     \issue{4}
% 	\pubyear{2007}

%     \title{\textbf{Qualitative Analysis of Forced Response of Blisks With Friction Ring Dampers}}
\begin{center}
   
   {\LARGE \textbf{Qualitative Analysis of Forced Response of Blisks With Friction Ring Dampers}}\\

   \vspace{1em}
   D. Laxalde$^{a,b}$, F. Thouverez$^{a}$, J.-J. Sinou$^{a}$, J.-P. Lombard$^{b}$
   
   \vspace{0.5em}
   ${}^{(a)}$ Laboratoire de Tribologie et Dynamique des Syst\`emes (UMR CNRS 5513)\\\'Ecole Centrale de Lyon, 36 avenue Guy de Collongue, 69134 Ecully Cedex, France\\
   ${}^{(b)}$ Snecma -- Safran group, 77550 Moissy-Cramayel, France
\end{center}
% 	\author[ECL,Snecma]{D. Laxalde\corauthref{cor1}}
% 	\corauth[cor1]{Corresponding author.}
% 	\ead{denis.laxalde@ec-lyon.fr}
% 	\author[ECL]{F. Thouverez}
% 	\ead{fabrice.thouverez@ec-lyon.fr}
% 	\author[ECL]{J.-J. Sinou}
% 	\ead{jean-jacques.sinou@ec-lyon.fr}
% 	\author[Snecma]{J.-P. Lombard}
% 	\ead{jean-pierre.lombard@snecma.fr}
	
% 	\address[ECL]{Ecole~Centrale~de~Lyon, Laboratoire~de~Tribologie~et~Dynamique~des Systemes (UMR CNRS 5513), \'Equipe~Dynamique~des~Structures~et~des~Systemes, 36,~avenue~Guy~de~Collongue, 69134~Ecully~Cedex, France}
% 	\address[Snecma]{Snecma~-~Safran~group, Rond~point~Rene~Ravaud~-~Reau, 77550~Moissy~-~Cramayel, France}
\begin{abstract}
	A damping strategy for blisks (integrally bladed disks) of turbomachinery involving a friction ring is investigated.
	These rings, located in grooves underside the wheel of the blisks, are held in contact by centrifugal loads and the energy is dissipated when relative motions between the ring and the disk occur.
	A representative lumped parameter model of the system is introduced and the steady-state nonlinear response is derived using a multi-harmonic balance method combined with an \emph{AFT} procedure where the friction force is calculated in the time domain.
	Numerical simulations are presented for several damper characteristics and several excitation configurations.
	From these results, the performance of this damping strategy is discussed and some design guidelines are given.
\end{abstract}
\section{Introduction}
Turbomachinery bladed disks operate in severe environments in terms of aerodynamic loads and may experience high vibratory stresses due to resonance or flutter. 
Their ability to withstand these stress levels, which may cause high cycle fatigue, mainly depends on the external inputs of damping. 
In bladed disk assemblies, friction in interfaces is the most widely used source of external damping; this includes shroud contact in blades or platform dampers. 
These devices were widely studied in the past and numerous methods of design and analysis are found in the literature.
The interest reader can find examples based on lumped parameter or equivalent models in works of \citet{Griffin-Review-friction}, \citet{Sinha-Friction_AeroUnstableRotor}, \citet{Ferri-Perturbation} or \citet{Csaba-JSV98} and, more recently, extended to more complex structures in works by \citet{Guillen-99}, \citet{Nacivet-JSV03} or \citet{Petrov-JT03} among others.
The contact interface could be one dimensional, two dimensional or three dimensional  \citep[e.g.][]{Sanliturk-JSV96,Chen-JSV00,Nacivet-JSV03}.

Beside, numerous experimental works were presented both in contact kinematic description (see \citet{Menq-Microslip2,Yang-JEGTP98-2} for example) and dynamic response prediction \citep[e.g.][]{Berthillier-FrictionDampers,Yang-JEGTP98-2,Sanliturk-Underplatform-Rainbow}.

However, in blisks (integrally bladed disks where disk and blades are a single piece), inherent damping is very low and mechanical joints no longer exist leading to the disappearance of nearly every sources of energy dissipation. 
Consequently, new methods and devices need to be found for the damping of blisks. 
Among others, friction damping using a ring is a solution which is found, experimentally, to be efficient on  single piece rotating structures (such as labyrinth seals for example). 
These rings are located in dedicated grooves underside wheels and contact is permanently maintained by centrifugal load due to engine rotation; friction and slipping occur during the differential motion of the two bodies in contact. 

As opposed to under-platform dampers in bladed disk assemblies, these devices were seldom studied; \citet{Ziegert-SplitRing} have presented methods of analysis and optimal design of split ring dampers for seals were the motion is studied using a quasi-static beam-like description (strength of materials).
This quasi-static description lead to a good understanding of the phenomenon and give some qualitative results. 
However, a nonlinear dynamics analysis is required to raise satisfying design rules.

In this paper, methods for predicting nonlinear steady-state response of blisks with friction ring dampers are presented along with some qualitative results based on a representative lumped parameter modelling.
The dynamical analysis is performed using the multi-Harmonic Balance Method ({\it HBM}). 
This method is found to be efficient and was widely used, in the past, to study friction dampers in bladed disks assemblies or shroud contacts. 
Some examples have been presented by \citet{Wang-HBM}, \citet{Sanliturk-JVA97} and \citet{Yang-JEGTP97} for mono-harmonic vibrations and by \citet{Pierre-Multiharmonic}, \citet{Petrov-JT03}, \citet{Nacivet-JSV03}, and \citet{Guillen-99} for multi-harmonic vibrations.

Results from numerical simulations (using this nonlinear analysis) are then presented and described.
Several parametric studies are presented in terms damper characteristics, contact characteristics, excitation configuration and mistuning are presented and some design guidelines are raised. 

\section{Lumped-parameter model of a blisk with friction ring damper}
Lumped parameter models have been widely used to study friction.
Beside from being simpler to study and compute (particularly for parametric studies), they can provide some interesting and valuable qualitative results if correctly designed.
In turbomachinery applications, several models have been proposed. Among others, we can mention \citet{Sinha-Friction_AeroUnstableRotor} or \citet{Wang-HBM} for friction damping applications and \citet{Griffin-Mistuning_Friction} and \citet{Lin-JSV96} for mistuning applications.

\begin{figure}[htbp]
\begin{center}
	\includegraphics[width=7.5cm]{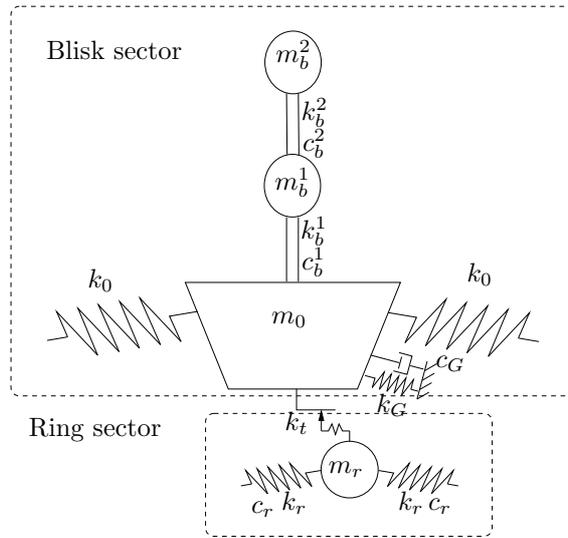}
	\caption{Lumped-parameter model}
	\label{fig:fig01}
\end{center}
\end{figure}

In this paper, we present a new lumped parameter model.
It consists in an elementary sector of a blisk and its associated friction ring element and it is represented in figure~\ref{fig:fig01}.

\begin{figure}[htbp]
\begin{center}
	\includegraphics[width=10cm]{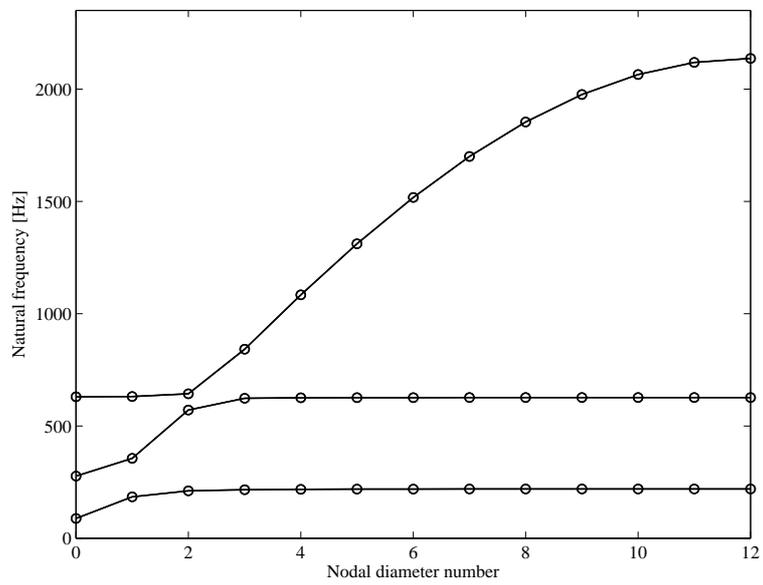}
	\caption{Frequency / diameter diagram}
	\label{fig:fig02}
\end{center}
\end{figure}

As this system is rotationally periodic (or cyclic), its modes are such that the deflection of a given elementary sector (described in figure~\ref{fig:fig01}) is the same as, but with a constant phase difference from, the preceding (or following) sector.
If $N$ is the number of elementary sectors of the structure, the phase angle is $2\pi n/N$ where $n$ is called the \emph{nodal diameter number}.
There are $N$ possible values of $n$ (from $0$ to $N-1$) but only $N/2+1$ values (from $0$ to $N/2$), if $N$ is even or $(N-1)/2+1$ values (from $0$ to $(N-1)/2$), if $N$ is odd, represent different mode shapes; the others being orthogonal to them with the same eigenfrequency.
Consequently, except for $n=0$ or $n=N/2$ if $N$ is even, the natural frequencies are repeated.
Further information on this subject can be found in the literature \citep[e.g.][]{Thomas-SC,Wildheim-JSV}.
As an example, figure~\ref{fig:fig02} displays the natural frequencies versus nodal diameter number of the (blisk) lumped parameter model with $N=24$ sectors.

The blisk lumped-parameter model was designed so that it stands as much as possible for a real blisk, particularly in terms of coupling between the blades and the disk.
There are three families of modes and regions of various modal density. 
This modal density is directly linked to the coupling between the blades and the disk in the particular mode.
A general rule, which for that matter applies to the present example, is that the blade/disk coupling can be important is low frequency/low nodal diameter and high frequency/high nodal diameter regions whereas in low frequency/high nodal diameter or high frequency/low nodal diameter regions this coupling is generally weak.
For example, in the two first families  the coupling between the blades and the disk is weak for high nodal diameters whereas in small nodal diameters the coupling is stronger.
Also note an important coupling appearing in the $2^{nd}$ and $3^{rd}$ modes with nodal diameter number 2.
Recalling that, in operating conditions, the external forcing generally acts on the blades this coupling parameter is an important factor regarding the efficiency of the friction ring damping since it determines the disk \emph{participation} in the global motion.
The energy dissipation due to friction occurs when the relative displacement between the disk and the ring is enough; this is achieved when the blisk is excited in areas of strong coupling where both the blades and the disk move.
The numerical values of the lumped-parameter are shown in table~\ref{tab:tab01}.

\begin{table}[htbp]
\centering
\caption{Numerical values of the model }
\begin{tabular}{cccccccc}
	$m_0$ & $k_0$ & $k_G$ & $m^1_b$ & $k^1_b$ & $m^2_b$ & $k^2_b$ & $\xi$ \\
	\hline
	1,2~kg & $5.10^7$~N/m & $6.10^5$~N/m & 350~g & $1.10^6$~N/m & 250~g & $2.10^6$~N/m & 1\textperthousand \\
	\hline
\end{tabular}
\label{tab:tab01}
\end{table}

Finally, concerning the ring, its mass and stiffness are chosen so that its lumped-parameter model represents a circular beam in extension motion and its stiffness $k_r$ represents the longitudinal rigidity of the ring.

\begin{figure}[htbp]
	\centering
	\subfigure[Local behaviour in the frictional region]{
		\includegraphics[width=7.5cm]{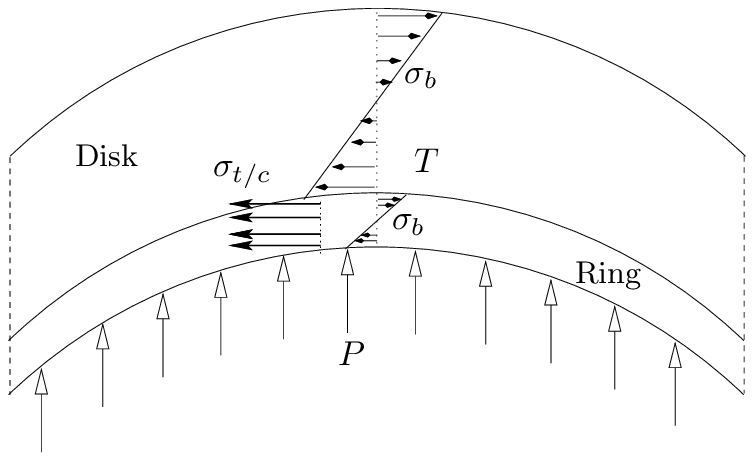}
		\label{fig:fig03a}}
	\subfigure[Friction ring in its groove]{
		\includegraphics[height=4cm]{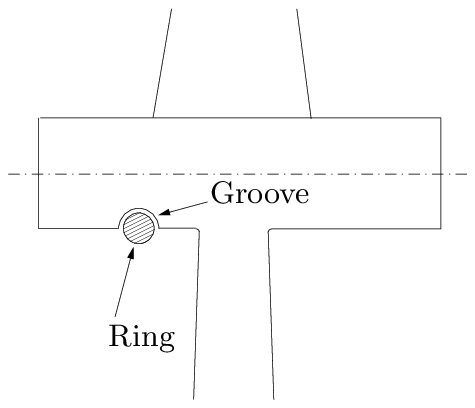}\label{fig:fig03b}}
	\caption{Physical model and local behaviour}
\end{figure}

An examination of the local behaviour of the friction area can help understanding the friction dissipation mechanism.
With reference to figure~\ref{fig:fig03a}, both the disk and the ring experience bending motion (bending stress $\sigma_b$) and are held in contact (slipping or not) due to the centrifugal load $P$. 
The friction forces in the interface ($T \leq \mu P$) between the two elements will generate tension or compression stresses ($\sigma_{t/c}$) mainly in the ring which ensure the continuity of stresses at the interface as shown in figure~\ref{fig:fig03a}. 
As a consequence the friction load is directly linked to the tension stresses and, assuming a linear elastic deformation, to the axial rigidity of the ring.

Figure~\ref{fig:fig03b} shows a physical model of the friction ring in its groove. 
In order to be assembled in these groove, underside the wheel of the disk, the ring needs to be split; the influence of this split on the dynamics of the system will be discussed later.

The system of equations of motion can be represented as:
\begin{equation} \label{eq:mvt}
		 \mat{M}\ddot{\vect{x}}(t)+\mat{C}\dot{\vect{x}}(t)+\mat{K}\vect{x}(t)=  	\vect{f}_{ext} (t) + \vect{f}_{nl} (t)
\end{equation}
where $\vect{x} = \left[ \vect{x}_d , \vect{x}_r , \vect{x}_{b_1} , \vect{x}_{b_2}  \right]^T$ is the physical displacements vector which gathers displacements of the disk, the blades and the ring degrees-of-freedom. 
The matrices $\mat{M}$, $\mat{C}$ and $\mat{K}$ are the structural matrices of mass, damping and stiffness respectively. 
The vector $\vect{f}_{ext}$ is an external distribution of rotating forces of Engine Order type, so that, considering a bladed disk with N sector, the $j^{th}$ sector's forcing term is:
\begin{equation}
	\vect{f}^{(j)}_{ext}(t)=\vect{f}^{(1)}_{ext} \exp{\left[i\left(\omega t - (j-1)\dfrac{2 \pi p}{N} \right) \right]},
	\label{eq:EOForce}
\end{equation}
where $p$ is engine order.

The vector of nonlinear forces $\vect{f}_{nl}$ due to friction is expressed, according to the Coulomb law, as:
\begin{equation}
	\vect{f}_{nl}(t)=-\mu P \mbox{sign}(\dot{\vect{u}}(t)),
	\label{eq:ForceCoulomb}
\end{equation}
where $\vect{u}(t)=\vect{x}_d(t)-\vect{x}_r(t)$ is the relative displacements vector between the ring and the disk, $\mu$ is the constant friction coefficient, and $P$ the normal contact load, which can be expressed for rotating structures as:
\begin{equation}
	P=m_r R \Omega^2
	\label{eq:NormalLoad}
\end{equation}
where $m_r$ is the ring's mass, $R$ is the distance  from the axe of rotation and the ring and $\Omega$ is the engine rotation speed.
Due to the high contact load resulting from the engine rotation, is assumed that the ring and blisk remain in contact at all times.

\section{Nonlinear Analysis}
\subsection{Harmonic Balance Method}
The aim of the nonlinear analysis performed here is to calculate the steady-state response of the system under a periodic external forcing. 
Among the variety of method available to achieve this goal, the Harmonic Balance Method \citep[][]{Nayfeh-Mook,Wanda} presents several advantages in terms of efficiency and accuracy with respect to its computational costs. 

The idea is to express the time-dependent variables of the nonlinear equation~(\ref{eq:mvt}) in terms of Fourier series (truncated in practice); the response $\vect{x}(t)$ is then:
\begin{equation}	\label{eq:RepFourier}
	\vect{x}(t)=\vect{X}^0 + \sum^{N_h}_{n=1} \{\vect{X}^{n,c} \cos (n\omega t) +  \vect{X}^{n,s} \sin (n\omega t)\}.
\end{equation}
The new unknowns of the problem are the Fourier components $\vect{X}^{n,c}$ and $\vect{X}^{n,s}$ which are balanced (using a Galerkin procedure) and the equation of motion can be rewritten using this formalism for each terms of equation~(\ref{eq:mvt}) in the frequency domain as:
\begin{equation}	\label{eq:MvtHBM}
	\mat{\Lambda} \vect{X} = \vect{F}_{ext} + \vect{F}_{nl} 
\end{equation}
with:
\begin{subequations}
	\begin{align}
		\vect{X}&=\left[\vect{X}^0,\vect{X}^{1,c},\vect{X}^{1,s},\dots,\vect{X}^{N_h,s}\right]^T,\\
		\vect{F}_{nl}&=\left[\vect{F}_{nl}^0,\vect{F}_{nl}^{1,c},\vect{F}_{nl}^{1,s},\dots,\vect{F}_{nl}^{N_h,s}\right]^T,\\
		\vect{F}_{ext}&=\left[\vect{F}_{ext}^0,\vect{F}_{ext}^{1,c},\vect{F}_{ext}^{1,s},\dots,\vect{F}_{ext}^{N_h,s}\right]^T.
	\end{align}
\end{subequations}

The matrix $\mat{\Lambda}$ is block-diagonal, $\mat{\Lambda}=diag\left(\mat{K},\mat{\Lambda}_1,\dots,
\mat{\Lambda}_{N_h} \right)$, and:
\begin{equation}	\label{eq:LambdaHBM}
\mat{\Lambda}_k=
\left( \begin{array}{cc}
-(k \omega )^2 \mat{M} + \mat{K} & k \omega \mat{C}           \\
-k \omega \mat{C}          & -(k \omega )^2 \mat{M} + \mat{K} \\
\end{array} \right)
\end{equation}

The problem~(\ref{eq:MvtHBM}) is nonlinear and is usually solved iteratively using a Newton-like method.

One of the issue of the {\it HBM} is that the Fourier components of the nonlinear forces $\vect{F}_{nl}$ can not, in most cases, be derived straightforward.
In particular, in problems involving friction, the nonlinear forces are only known (using equation~(\ref{eq:ForceCoulomb})) in the time domain and the expansion in the frequency domain is not obvious.
When few harmonics (no more than $3$ in practice) are retained an analytical derivation of these frequency domain components is possible \citep[e.g.][]{Nayfeh-Mook,Wang-HBM}.
To overcome this problem when more harmonics are considered, the use of an {\it Alternating Frequency Time} method \citep[e.g.][]{Cameron-AFT} can be a solution.
This method will be detailed in the next section.

\subsection{AFT Method}
The aim is, in the Newton procedure, to derive the Fourier components of the nonlinear forces as a function of the displacements Fourier components.
In computational applications, the idea of the {\it AFT} method is to use Discrete Fourier Transformation (DFT) to derive the Fourier components of the nonlinear forces for given displacements in the frequency domain.

Let's start from $\vect{X}^{n,c}$ and $\vect{X}^{n,s}$, the harmonic components of the response, predicted by a given iteration of the Newton-like method.
Using, an IDFT ({\it Inverse Discrete Fourier Transform}) procedure, one can express the associated  displacements $\vect{x}(t)$ and velocities $\dot{\vect{x}} (t)$ in the time-domain.
Then using a nonlinear operator the nonlinear force in time-domain is derived (here this step is achieved using equation~(\ref{eq:ForceCoulomb}) as described in the next section). 
Finally, using a DFT ({\it Discrete Fourier Transform}) algorithm one expresses the harmonic components of the nonlinear force in the frequency-domain. 
The following diagram illustrates this procedure:
$$
\begin{CD}
	\vect{X}^{n,c}, \; \vect{X}^{n,s}  @>{\operatorname{IDFT}}>>  \vect{x}(t), \; 
	\dot{\vect{x}} (t) \\
	@.      @VVV \\
	\vect{F}_{nl}^{n,c}, \; \vect{F}_{nl}^{n,s}    @<<{\operatorname{DFT}}< 
	\vect{f}_{nl}(\vect{x},\dot{\vect{x}},t)
\end{CD}
$$

\subsection{Calculation of friction forces in the time domain}
At a given iteration of the Newton scheme and during the {\it AFT} procedure, the friction forces need to calculated, in the time domain, in accordance with the velocity and displacement given (by the Newton method). 
This is generally done using the Coulomb law or, more usually, using a regularized Coulomb law and several strategies can be used.
Some authors have proposed a regularized expression of the sign function \citep[e.g.][]{Berthillier-FrictionDampers} or arctangent or hyperbolic tangent function \citep[e.g.][]{Petrov-ISMA02}, some others have used the theory of distributions \citep[e.g.][]{Pierre-Multiharmonic}, and some have used an iterative procedure in the time domain \citep[e.g.][]{Poudou-AIAA03,Guillen-99,Petrov-JT03}.
For the latter strategy, it is convenient to add a tangent stiffness to each friction element; doing so, the Coulomb law can be used straightforward to compute at each time step the equilibrium of the contact point.
In previous studies on under-platform dampers, this stiffness was included in the damper model.
However, in general contact problem, this tangent stiffness appears as an equivalent penalty stiffness and can then account for the elasto-plastic shear deformations of the contact asperities of the bodies in contact.

\begin{figure}[htbp]
	\centering
	\includegraphics{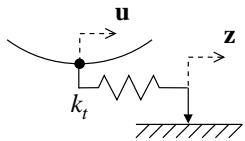}
	\caption{Friction element (relative displacements)}
	\label{fig:fig04}
\end{figure}

Figure~\ref{fig:fig04} shows a friction element defined in relative displacements, with a tangent stiffness:
\begin{itemize}
	\item $\vect{u}$ is the relative between two points in contact, $\vect{u}=\vect{x}_d-\vect{x}_r$ in our example.
	\item $\vect{z}$ is the relative displacement of the contact point.
\end{itemize}
The nonlinear forces vector given by equation~(\ref{eq:ForceCoulomb}) becomes:
\begin{equation} \label{eq:force_frottement_tempo}
	\vect{f}_{nl}(t)=
	\begin{cases}
		-k_t \left(\vect{u}(t)-\vect{z}(t)\right) & \mbox{\  if \ }  k_t |\vect{u}(t)-\vect{z}(t)|\leq \mu P, \\
		-\mu P \mbox{sign} \left(\dot{\vect{z}}(t)\right) & \mbox{\  if \ } k_t |\vect{u}(t) -\vect{z}(t)|\geq \mu P.
	\end{cases}
\end{equation}

In the time domain of the {\it AFT} procedure, velocities and displacements are given in one period of the motion.
These are entries for the computation of the associated nonlinear forces in one period of the motion.
This computation has to be performed time-iteratively (like a numerical integration) because of the history dependency of the friction law.
For example, given the relative displacements and velocities at time $t_i$, $\vect{u}(t_i)$ and $\dot{\vect{u}}(t_i)$ and the displacement of the contact point $\vect{z}(t_i)$, the friction force can be predicted (using equation~(\ref{eq:force_frottement_tempo})) for time $t_i$  assuming a sticking state ($\vect{z}(t_i)=\vect{z}(t_{i-1})$) for the contact point:
\begin{equation}	\label{eq:Force_fr_pre}
	\vect{f}_{nl}^{pre}=-k_t \left(\vect{u}(t_i)-\vect{z}(t_{i-1})\right).
\end{equation}
The Coulomb law is then used to correct the friction force and the displacement of the contact point:
\begin{subequations}
\begin{align}
	\vect{f}_{nl}(t_i)&=
	\begin{cases}
		\vect{f}_{nl}^{pre} & \mbox{ if } |\vect{f}_{nl}^{pre}| < \mu P \\
		-\mu P \cdot \dfrac{\vect{f}_{nl}^{pre}}{|\vect{f}_{nl}^{pre}|} & \mbox{ if } |\vect{f}_{nl}^{pre}| \geq \mu P,
	\end{cases}\\
	\vect{z}(t_i)&= \vect{u}(t_i) -\dfrac{\vect{f}_{nl}(t_i)}{k_t}.
\end{align}
\end{subequations}
Using this iterative scheme, the periodic friction force can be derived.

\section{Numerical results}
In this section, several aspects of friction ring damping are investigated through numerical simulations and discussed.
All nonlinear numerical simulations were performed using the multi-harmonic balance procedure and a third order Lagrange predictor was used for the continuation of the frequency responses.
A $d^{th}$ order Lagrange predictor, see \citet{Stoer-NumericalAnalysis} for example, requires the last $d+1$ points to predict the next one $\hat{\vect{X}}^{p+1}$ and an interpolation is build on the arc-length $s$ (which is here the excitation frequency) as:
\begin{equation}
	\hat{\vect{X}}^{p+1}=\sum_{i=0}^d L_i \cdot \vect{X}^{p-d+i} \mbox{ with } L_i=\prod_{\begin{array}{c}\\[-0.95cm]
		\scriptstyle j=0\\[-0.4cm]
		\scriptstyle j\neq 1
	\end{array}}^d \dfrac{s^{(p+1)}-s^{(p-d+j)}}{s^{(p-d+i)}-s^{(p-d+j)}}.
\end{equation}
The step $\Delta s$ is adjusted {\it a posteriori} as a function of the (Newton-like) corrector's performance (number of iterations) at the previous step.

All frequency response plots represent the blade response (degree-of-freedom $x_{b_2}$ in figure \ref{fig:fig01}), with, in dashed line, the linear response (without ring) and in solid lines, the nonlinear responses (with friction ring).
Parametric studies on the type of excitation (or blade/disk coupling as mentioned earlier), on contact parameters, damper characteristics are presented and the efficiency of this damping strategy is discussed.
Then the influence of ring's split discussed.
And finally, the impact mistuning and friction ring damping is investigated.

\subsection{Optimal damper mass -- Blade/disk coupling}
In this first part, the influences of the ring damper mass and of the excitation configuration are investigated.
The rotation speed is kept constant is these simulations; however, note that as the ring's mass changes, the contact normal load also change with respect to equation~(\ref{eq:NormalLoad}).

\paragraph*{Strongly blade/disk coupling}

With reference to the frequency/nodal diameter map of figure~\ref{fig:fig02}, we first focus on a small engine order excitation ($p=2$ for example in equation~(\ref{eq:EOForce})) with frequency range around the first modal family (similar behaviour would appears in other modal family however).

\begin{figure}[htpb]
	\begin{center}
		\includegraphics[width=.7\textwidth]{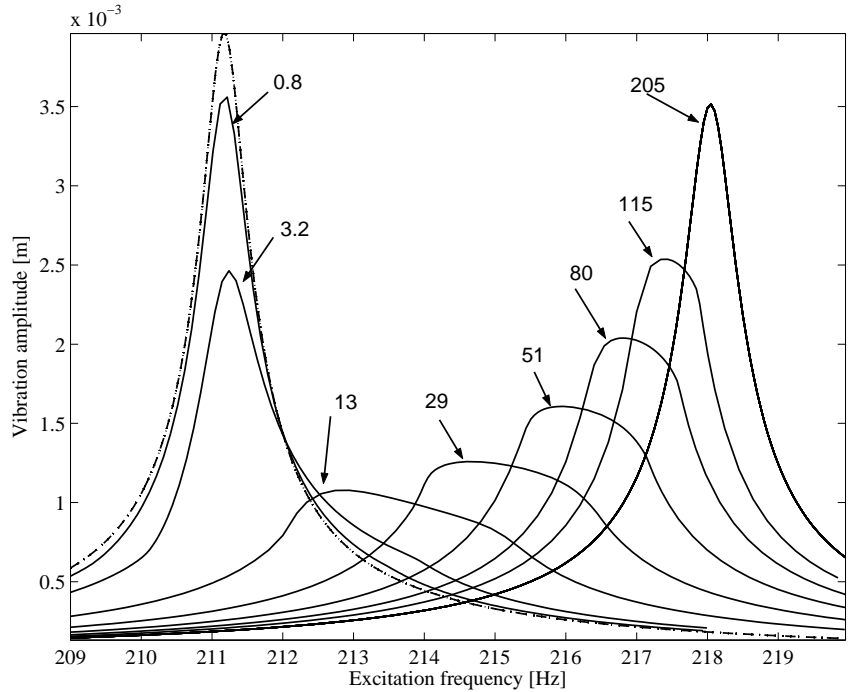}
	\end{center}
	\caption[Frequency responses with strong blade/disk coupling]{Frequency responses with strong blade/disk coupling; (- - -)~: linear response, (-----)~: nonlinear responses at different damper mass (g)}	
	\label{fig:fig05}
\end{figure}

Several frequency responses are plotted in figure~\ref{fig:fig05} for different damper mass.
From these response, we can first notice that the damping is efficient in this configuration when the ring is sufficiently thick and that an optimal value appears ($m_{ring}\approx 20g$).
At this optimal value, the reduction of vibration amplitude is nearly 4.
Also note the progressive frequency shift of the resonance peaks as the ring's mass increase, this is due to two facts combined.
First, considering the associated linear problem, as the ring's thickness increases the global system becomes stiffer which account for the frequency shift to the higher frequency.
And second, the nonlinear effect of friction has usually a softening effect on system's dynamic and, with respect to figure~\ref{fig:fig05}, as the ring's mass decreases the contact normal load also decreases (assuming a fixed rotation speed) and more and more slipping occurs which account for the frequency shift to the lower frequency as the damper's mass decreases.

\paragraph*{Weakly blade/disk coupling}

\begin{figure}[htpb]
	\begin{center}
		\includegraphics[width=.7\textwidth]{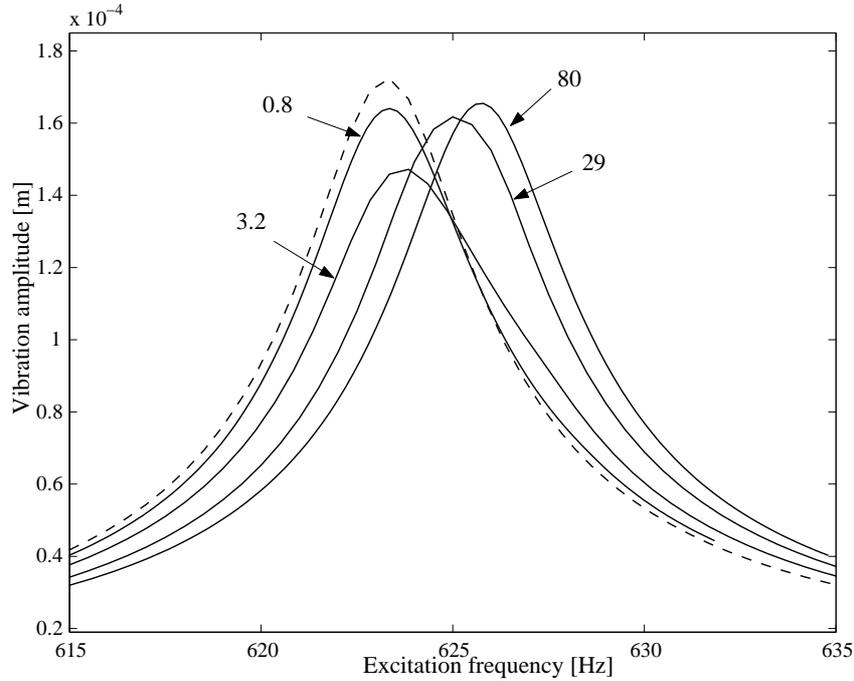}
	\end{center}
	\caption[Frequency responses with weak blade/disk coupling]{Frequency responses with weak blade/disk coupling;  (- - -)~: linear response, (-----)~: nonlinear responses at different damper mass (g)}	
	\label{fig:fig06}
\end{figure}

A weaker blade/disk coupling can be found in higher nodal diameter in figure~\ref{fig:fig02}.
As an example, let's consider a 6 engine order excitation on the second modal family and the associated frequency responses plotted in figure~\ref{fig:fig06}.
In this example, the friction ring damper is clearly less efficient; even if an optimal mass appears (around 5g), the reduction of vibration's level is not as important as in strong coupling configuration.
In effect, in the configuration, there is not enough motion in the disk as the blades are excited and as a consequence no relative motion between the disk and the ring can occur.

\subsection{Friction ring damping efficiency at different operating speeds}
In this section, we focus on the influence of the normal contact load (which is, in our application, related to the engine rotation speed) on the friction ring damping efficiency.
To do so, the damper mass is fixed as well as the excitation parameters (Engine Order 2) and several simulations were performed for different values of the engine rotation speed.

\begin{figure}[htpb]
	\begin{center}
		\includegraphics[width=0.7\textwidth]{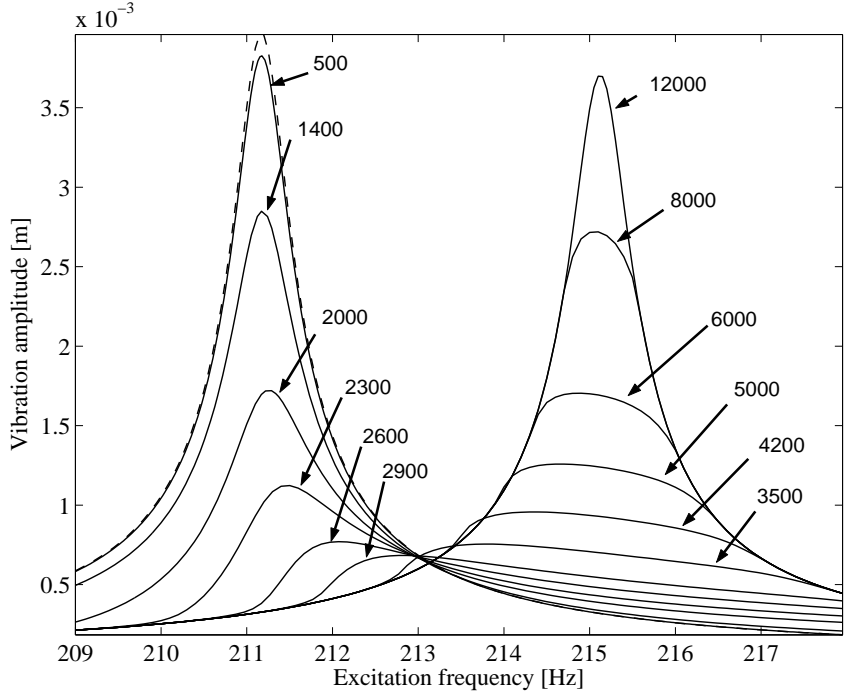}
	\end{center}
	\caption[Frequency responses at various rotation speed]{Frequency responses at various rotation speed; (- - -)~: linear response, (-----)~: nonlinear responses at different rotation speeds (rpm)}
	\label{fig:fig07}
\end{figure}
In figure~\ref{fig:fig07}, the results are depicted and the first remarks is that the existence of an optimal rotation speed/normal load is clear around 3000 rpm.
Moreover, one can see that that in a quite large vicinity of this optimal value (from 2000 to 6000 rpm), the friction ring damping is still efficient.

This is a key feature since a turbomachinery often operates at different rotation speeds (take-off, cruise, landing, \dots).
As a consequence, the rings can, for example, be optimally designed for the cruise speed and be also efficient at take-off or landing speeds.

\subsection{Symmetry breaking induced by ring's split}
\begin{figure}[htbp]
\centering
	\includegraphics[width=8.5cm]{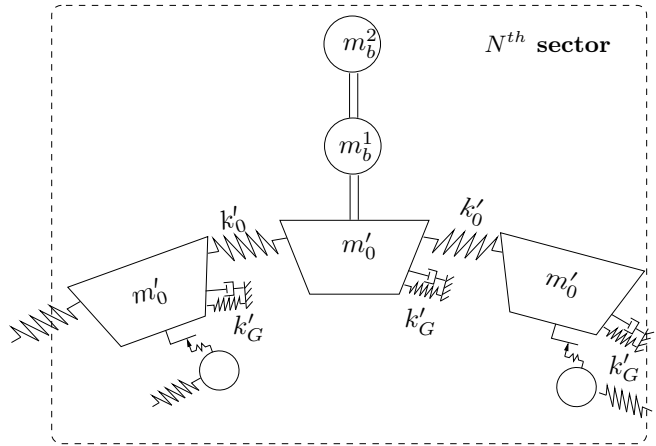}
	\caption{Fine lumped-parameter model}
	\label{fig:fig08}
\end{figure}

As mentioned earlier, in order to be assembled the ring need to be split.
However, considering the model described in figure~\ref{fig:fig01}, a split in the ring would necessary spread over one elementary sector and generate an exaggerated gap compared to the technological reality.
As a consequence, we built a refined model, presented in figure~\ref{fig:fig08}, in which the disk mass, $m_0$, is refined into three masses, $m_0^{\prime}$.
This allows the split to spread only over a portion of one sector.

\begin{figure}[htbp]
	\centering
	\subfigure[Continuous ring]{
	\includegraphics[width=.47\textwidth]{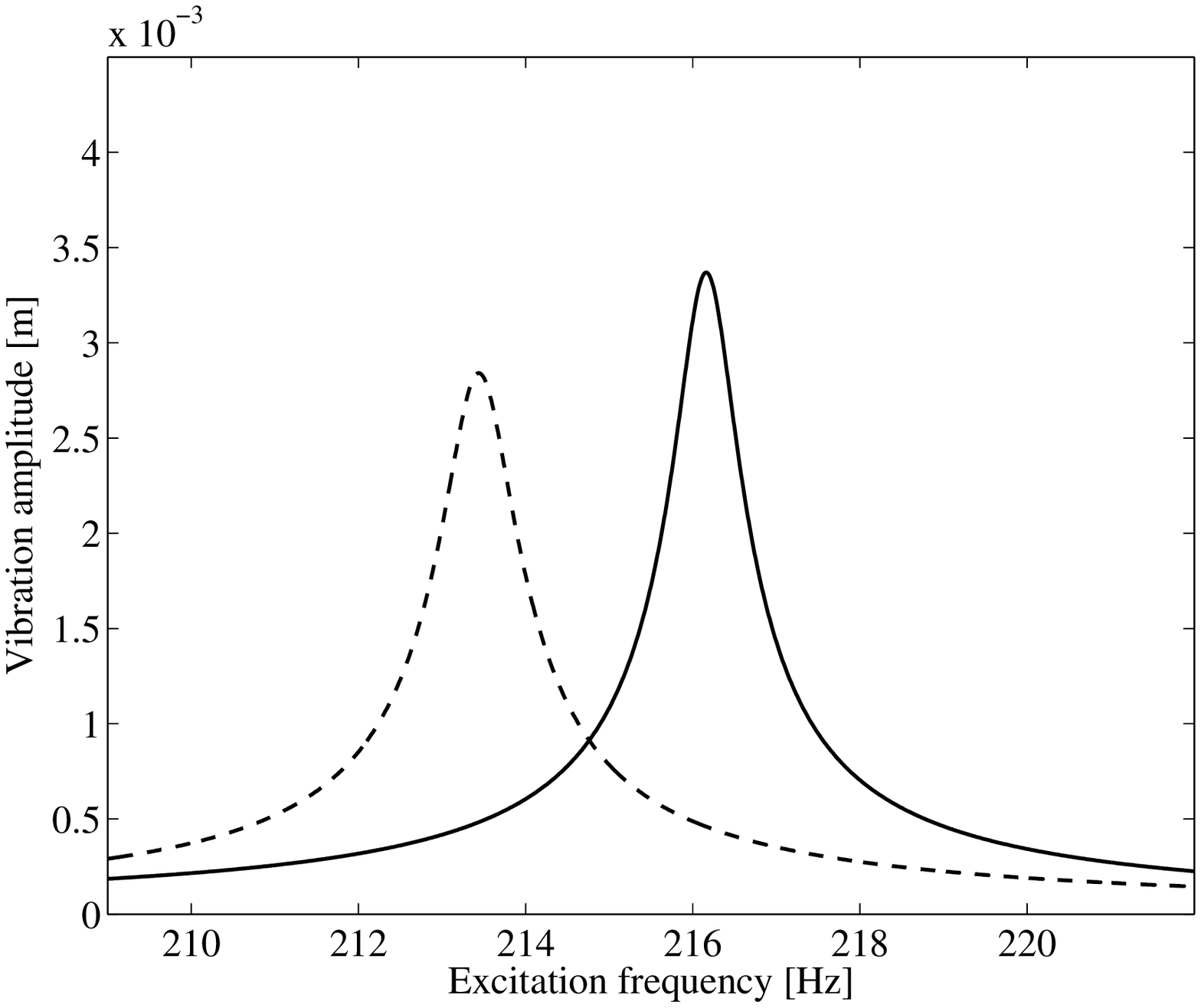}
	\label{fig:fig09a}}
	\subfigure[Split ring]{
	\includegraphics[width=.47\textwidth]{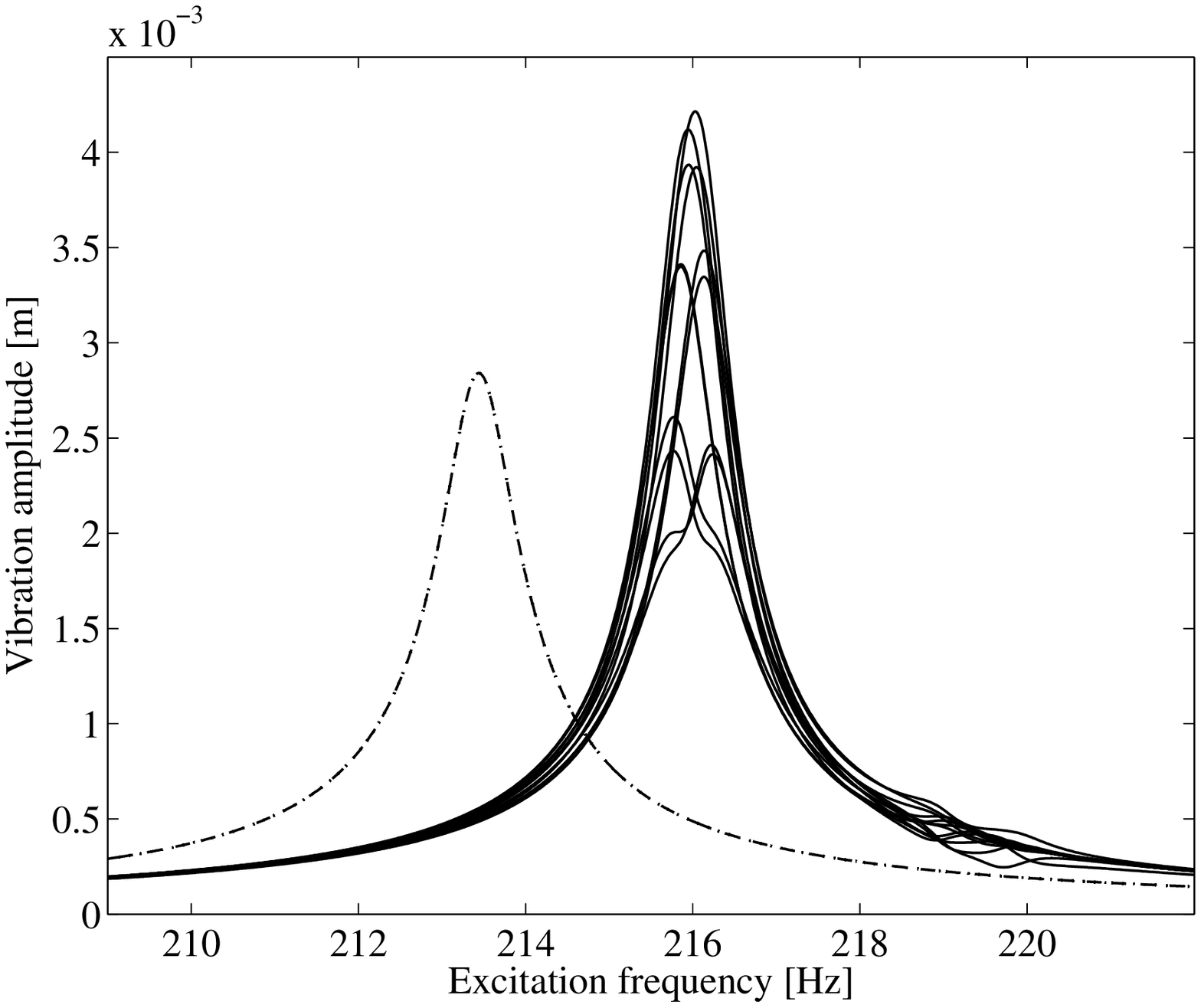}
	\label{fig:fig09b}}
	\caption[Linear frequency responses with and without ring's split]{Linear frequency responses with and without ring's split; (- - -)~: without a ring, (-----)~: with ring (split or not).}
	\label{fig:IfluCoup}
\end{figure}

The influence of the ring's split was then investigated by comparing the dynamic response of the system with or without a split.
First, the ring is assumed to be fully stuck (no sliding occurs) and the system is then linear.
The frequency response plots resulting from these simulations are displayed in figure \ref{fig:IfluCoup}.

We focused on the first resonance peak of the 2 engine order excitation and comparing the blade's resonant responses with (figure~\ref{fig:fig09b}) or without (figure~\ref{fig:fig09a}) ring's split, it appears that each blades have a different response and that the maximum level is more important when the ring is split.
This symmetry breaking, which corresponds to the separation of the initial twin modes (two-fold degeneracy), is similar to the consequence of mistuning and can lead to localization phenomena (see \cite{Pierre-Localization1,Pierre-Localization2} for example).

\begin{figure}[htpb]
	\begin{center}
		\subfigure[Continuous ring]{\includegraphics[width=.47\textwidth]{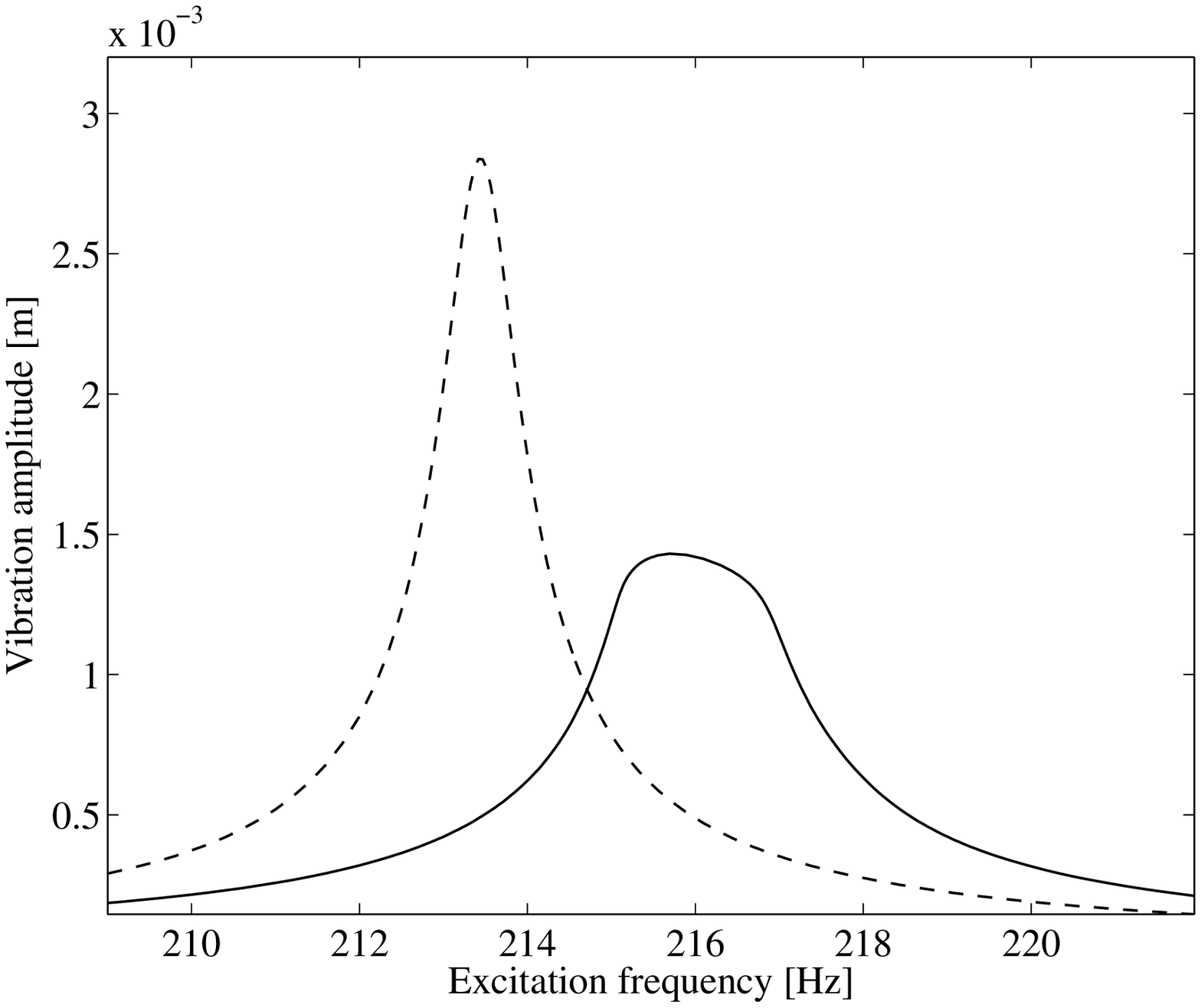}
		\label{fig:fig10a}}
		\subfigure[Split ring]{\includegraphics[width=.47\textwidth]{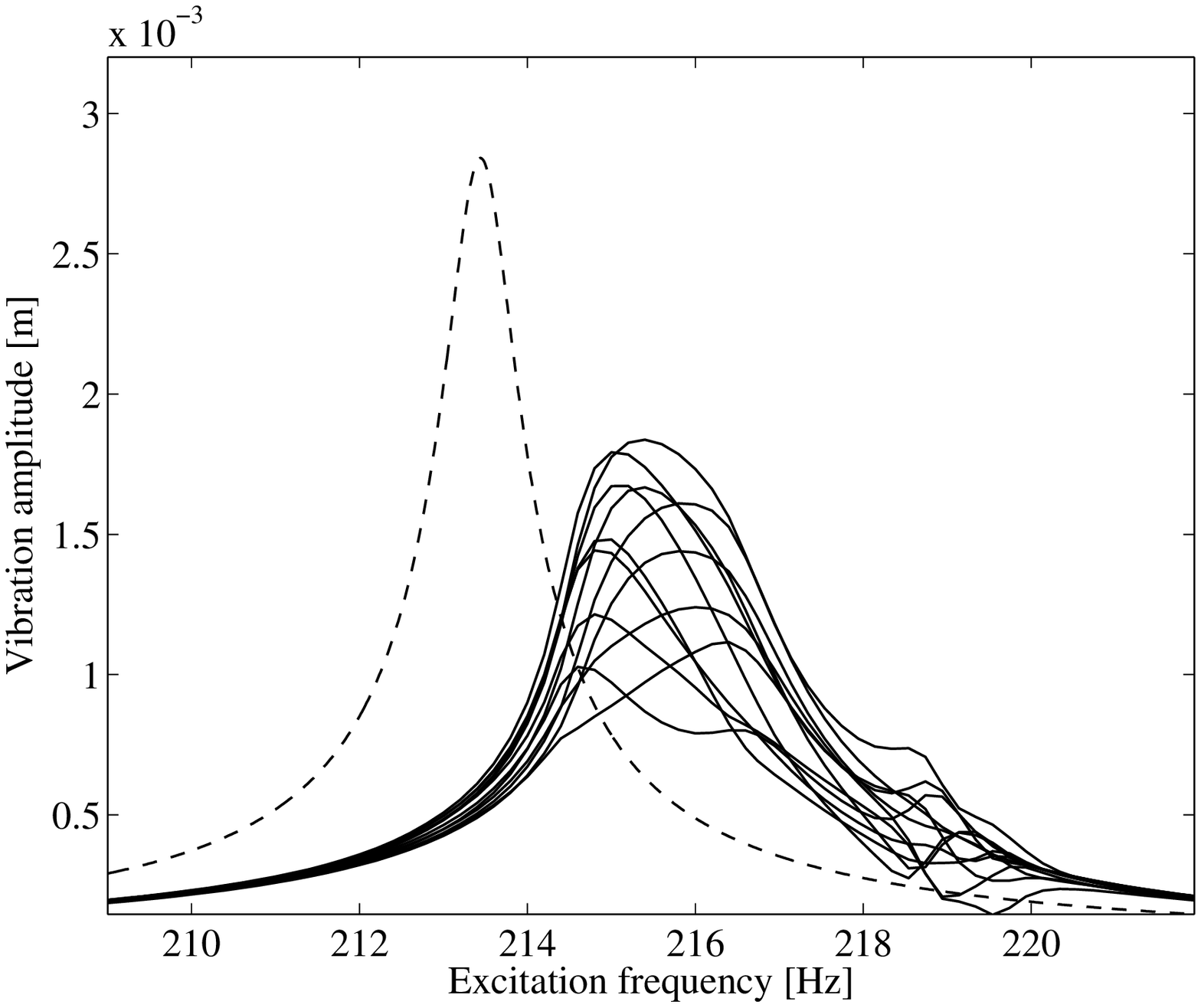}
		\label{fig:fig10b}}
	\end{center}
	\caption[Nonlinear frequency responses with and without ring's split]{Nonlinear frequency responses with and without ring's split; (- - -)~: without a ring, (-----)~: nonlinear responses.}
	\label{fig:InfluCoupNL}
\end{figure}

When the nonlinear effect are introduced (figure~\ref{fig:InfluCoupNL}), the same phenomena of symmetry breaking and increasing of resonant responses appear.
However, the response amplification is limited by the nonlinear effect and then smaller than in the linear case.

These results show the influence of the split in the ring damper on the global dynamics of a bladed disks.
However, these observations have to be put in perspective with respect to the influence of the inherent mistuning (discussed next) of a real bladed disks which may generally be more important.

\subsection{Influence of mistuning}
To conclude on the numerical results, the influence of mistuning was investigated.
Mistuning refers to small variations in the properties of a system originally designed to be cyclically symmetric. 
These variations are mainly due to manufacturing tolerances or material inhomogeneity and are inevitable. 
Mistuning can lead to significant changes in dynamical behaviour with respect to the tuned case as presented by \citet{Pierre-Localization1,Pierre-Localization2}; vibratory amplitudes in forced response are increased and impact on high cycle fatigue can be considerable. 
These effects are even more important in Integrally Bladed Disks since their "natural" damping is far smaller than for bladed disks assemblies. 
The influence of mistuning on friction damping has been studied earlier by \citet{Griffin-Mistuning_Friction} or \citet{Lin-JSV96} and the performance of the damping technology discussed here as applied to blisks may be influenced by this phenomenon. 

Considering the lumped parameter model of figure \ref{fig:fig01}, mistuning was introduced as random perturbations in the blades' stiffness $k^1_b$ and  $k^2_b$ as it is often the case since the disk is assumed to be symmetric. 
These perturbations lead to changes in the natural frequencies of the blades (cantilevered).

\begin{figure}[htpb]
	\begin{center}
		\subfigure[Tuned bladed disk]{
		\includegraphics[width=.47\textwidth]{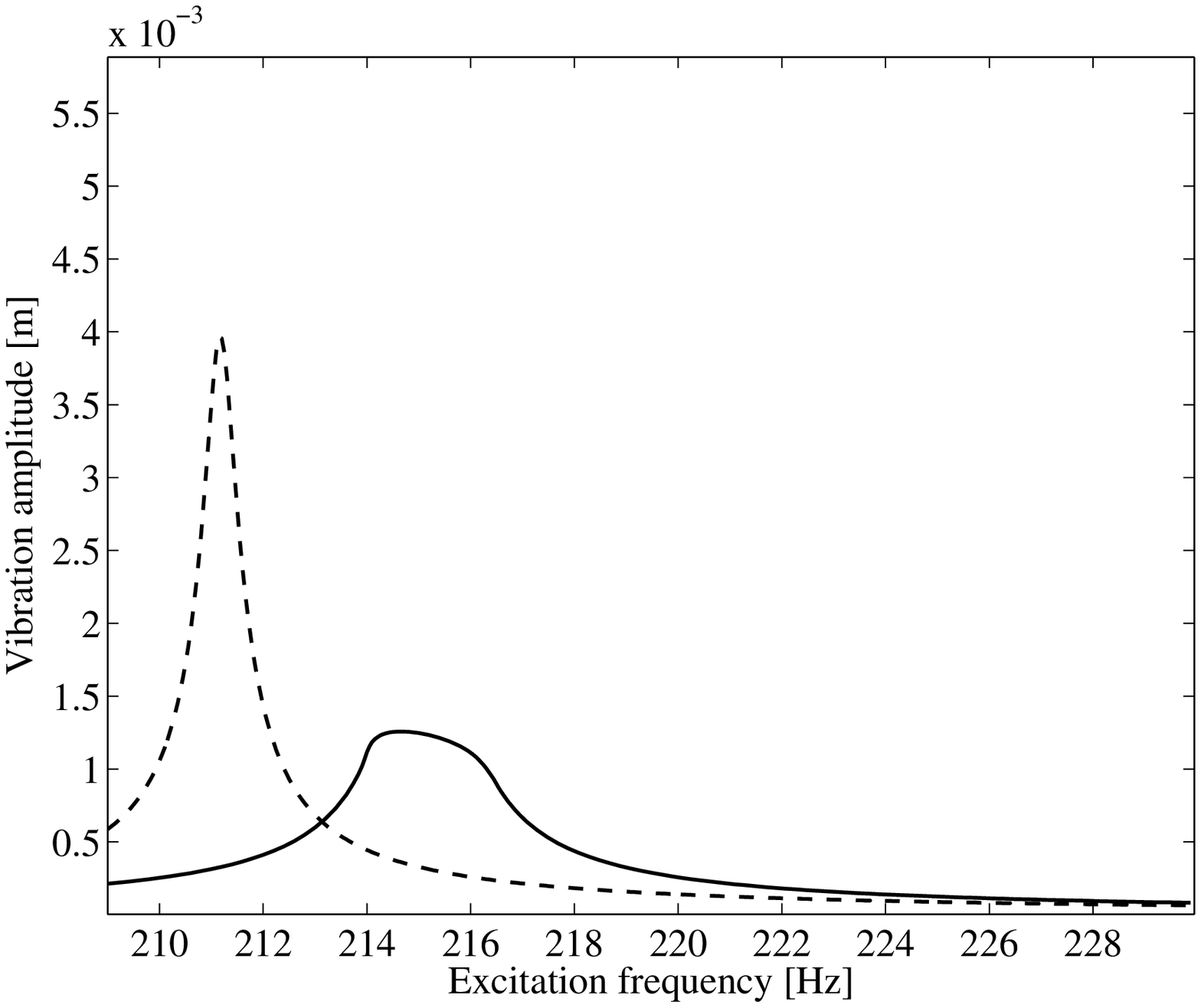}
		\label{fig:fig11a}}
		\subfigure[Mistuned bladed disk $\sigma=1.2$\%]{
		\includegraphics[width=.47\textwidth]{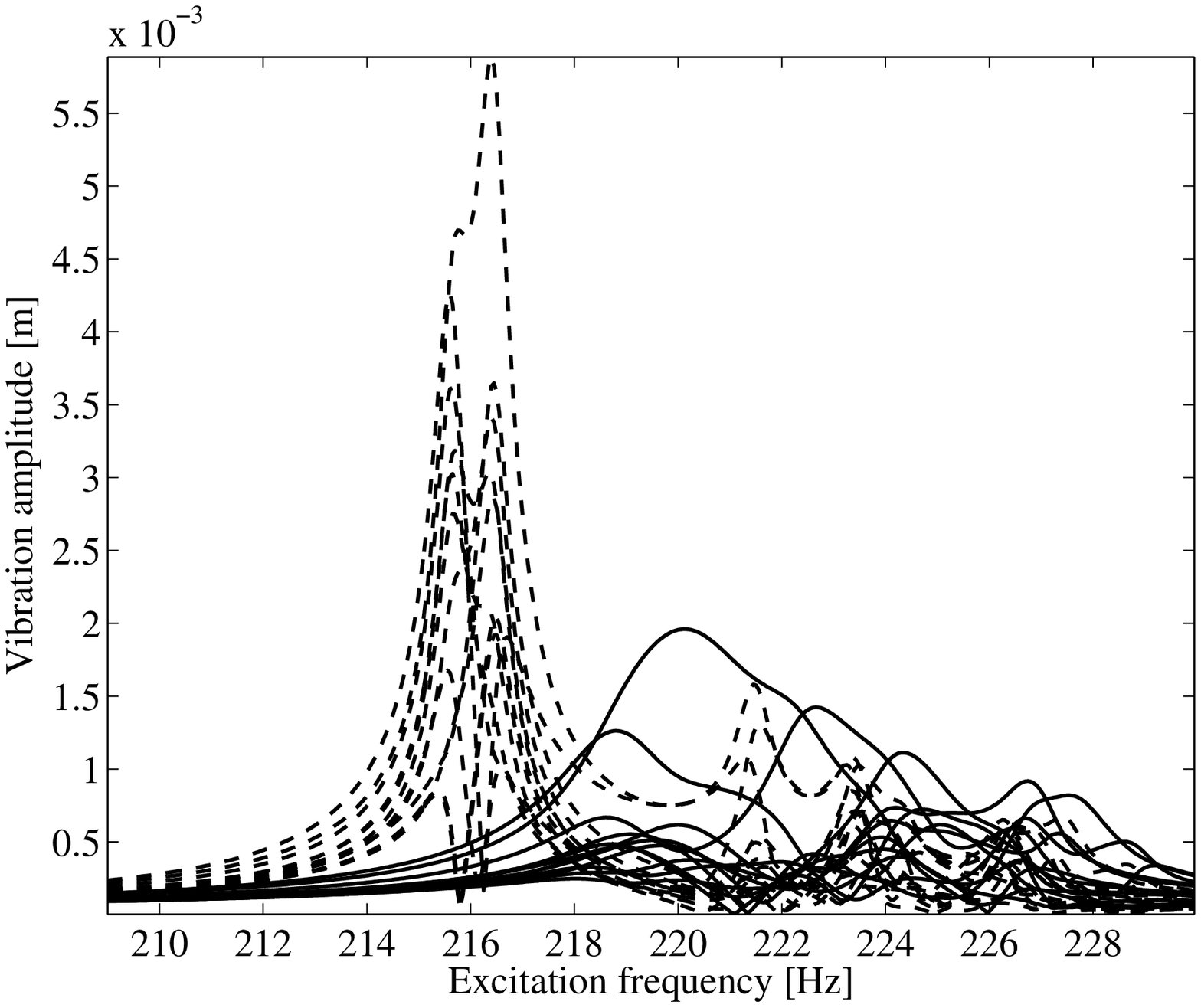}
		\label{fig:fig11b}}
	\end{center}
	\caption[Frequency responses with and without mistuning]{Frequency responses with and without mistuning; (- - -)~: linear responses, (-----)~: nonlinear responses.}
	\label{fig:fig11}
\end{figure}

A frequency response is shown in figure \ref{fig:fig11}, the mistuning standard deviation is about 1.2\% of the cantilevered blade's nominal natural frequency. 
The resonant responses of the blades are displayed for the tuned and the mistuned case with or without ring.
First, we can notice the typical feature of mistuned responses which is the break of symmetry which leads to a separation of the resonance peak (with or without rim damping).
Then, the vibratory amplitudes are increased in the mistuned case with respect to the tuned case; however, the damping is still efficient in presence of mistuning.

\subsection{Conclusions on the friction ring damping performance}
The results of these numerical simulations clearly highlight the possibilities and limits of the friction ring damping strategy. 
In essence, the dissipation mechanism of friction is efficient only if some relative motion between the two bodies (the disk and the ring) in contact occurs. 
As opposed to under-platform dampers, the rings are not directly in contact with the blades to which the external forcing applies. 
As a consequence, the coupling between the blades and the disk which conditions the energy transfer from the blades to the disk is an essential criterion. 
It has been demonstrated that when the coupling is weak (that is when little energy get transferred to the disk), the ring damping is less efficient.
On the other hand, when this coupling is sufficiently strong the rings are efficient. 
Finally, it was shown that the contact parameters (normal loads or friction coefficient) are also significant in the friction damping performance.
The qualitative results given by these simulations are quite similar to those observed with under-platform damper; the alternation between slipping and sticking states results in frequency shift which offsets the resonant frequency from its initial value ({\it i.e.} non damped), and an optimum can be reached for a given excitation configuration in the ring's mass.

\section{Conclusions}
A strategy of damping for Integrally Bladed Disks (blisks) using a friction split ring was investigated. 
A procedure of analysis using a multi-Harmonic Balance Method and an {\it AFT} method was described and applied for calculating the steady-state nonlinear response of a system under periodic excitation.

Simulations on the lumped-parameter model presented lead to qualitative results in terms of energy dissipation with respect to several configurations.
Some guidelines on the use of this technology, its limits and performances, were highlighted.
We show that the coupling ratio between the disk and the blades was a critical parameter of the damping power of such rings.

\section*{Acknowledgment}
Thanks go to Snecma for their technical and financial supports.

\end{document}